# The Plastic Origin of van der Waals material GaGeTe

Qiao Wang, Ping-An Hu *

Key Laboratory of Micro-systems and Micro-structures, Manufacturing of Ministry of Education
Harbin Institute of Technology
Harbin 150001, China

Plastic inorganic semiconductors are promising candidates for high-performance stable flexible electronics. Germanium-based chalcogenide materials are well-known for excellent semiconducting properties, specifically superior carrier mobility. However, these materials typically exhibit inherent brittleness, which in germanium tellurides originates from the intrinsic half-bond nature of the metavalent Ge–Te bond. Here, we report an ultra-high plasticity germanium-based chalcogenide of GaGeTe. Bulk GaGeTe can sustain a large engineering strain up to 7.5% without fracture. Through a comprehensive multi-scale investigation ranging from macroscopic characterization to atomic-resolution scanning transmission electron microscopy, we reveal that the exceptional plasticity is due to a synergistic effect of interlayer gliding and intralayer lattice distortion. DFT calculations further elucidate the chemical origin. The existence of Ga atoms avoids the formation of brittle metavalent Ge-Te bonds. Instead, a robust honeycomb lattice formed by strong Ga-Te, Ge-Ge, and Ga-Ge bonds preserves high bond strength under large tensile strains in simulations, thereby underpinning the macroscopic plastic behavior. This systematic study on the origin of the plasticity in GaGeTe introduces bond engineering as a potential approach for constructing plastic inorganic semiconductors.

## 1. Introduction

The rapid evolution of flexible electronics has created an urgent demand for materials that possess a comprehensive suite of merits: mechanical flexibility, physicochemical stability, and high electron mobility. However, conventional material systems struggle to meet these requirements simultaneously.[1–3] Traditional inorganic semiconductors exhibit excellent electrical properties but are inherently brittle, restricting their direct integration into flexible devices.[1] Though they can be fabricated into polycrystalline thin films to accommodate deformation, these films suffer from grain boundary defects compared to their single-crystal counterparts, which severely compromise carrier transport, thereby degrading detection performance.[4] While organic semiconductors offer intrinsic mechanical compliance, they are plagued by insufficient carrier mobility and poor environmental stability.[3,5–7] Most organic semiconductors are particularly susceptible to degradation under ultraviolet (UV) irradiation and high temperatures,[6,7] while prone to embrittlement at low temperatures,[8] resulting in a limited operational lifespan and inconsistent performance in harsh conditions. Plastic inorganic semiconductors have emerged as a transformative solution to this contradiction. By combining the high carrier mobility, thermal stability, and UV resistance of inorganic materials with the mechanical flexibility typically associated with organics, they meet the rigorous performance criteria required for next-generation electronics.[9] Consequently, plastic inorganic semiconductors hold immense promise for critical applications in flexible and wearable technologies, including thermoelectric devices[10], touchscreens[11], X-ray detectors[12], and ferromagnetic semiconductors[13].

Although progress has been made in plastic inorganic semiconductors,[10] plastic materials that are based on germanium (Ge) remain exceptionally rare. This gap draws particular attention since Ge-based semiconductors are renowned for their superior carrier mobility and excellent optoelectronic properties, features that are critical for high-speed logic circuits[14,15], advanced optoelectronic systems[16,17], efficient thermoelectrics and cutting-edge phase-change memory devices. The conventional binary Ge-chalcogenides (sulfides, selenides, and tellurides) are inherently brittle.[18] In the case of tellurides (GeTe), this failure is primarily attributed to the dominance of metavalent bonds featuring a half-bond order between Ge atoms and Te atoms that easily succumb to fracture under stress. To overcome this fundamental limitation, we devised a targeted bond engineering strategy exemplified by the synthesis of the multicomponent single crystal, GaGeTe. Its fundamental building block is a covalently bonded sextuple layer characterized by a honeycomb structure with a specific Te-Ga-Ge-Ge-Ga-Te stacking order, in which a chair-like germanene-like Ge layer is sandwiched between two GaSe-type GaTe outer slabs.[19] In this ternary architecture, the incorporation of Gallium (Ga) is critical; it effectively suppresses the formation of brittle Ge–Te bonds, replacing them with a robust network composed of distorted $sp^3$-hybridized Ga-Te, Ge-Ge, and Ga-Ge bonds. These bonds maintain high strength even under significant strain, endowing the material with macroscopic plastic behavior and marking the realization of the first Ge-based plastic inorganic semiconductor.

Furthermore, the understanding of the deformation mechanisms in plastic van der Waals materials remains elusive. Existing studies predominantly attribute the plasticity of these materials to classical mechanisms such as dislocation[20,21], phase transitions[22], twinning[23], and amorphization[24,25]. While these investigations provide valuable insights into defect dynamics and structural transformations, they largely overlook critical intralayer contributions, particularly intralayer lattice distortion, leaving the deeper, finer-scale structural responses underexplored. To address this deficiency, we performed a comprehensive atomic-scale investigation of GaGeTe using spherical aberration-corrected transmission electron microscopy (AC-TEM) coupled with density functional theory (DFT) calculations. Our study reveals a unique deformation mechanism that distinguishes GaGeTe from previously reported materials: the plasticity in this van der Waals inorganic semiconductor is co-dominated by interlayer gliding and intralayer lattice distortion. This finding not only elucidates the origin of the ultra-high plasticity in Ge-based material GaGeTe but also establishes a more complete theoretical framework for understanding the deformation mechanisms of plastic inorganic semiconductors.

## 2. Results and Discussion

As an emerging ternary van der Waals (vdW) semiconductor, GaGeTe crystallizes in a hexagonal crystal lattice (space group $R\bar{3}m$, lattice constants a = 4.05 Å and c = 34.65 Å; **Figure 1**a-I and Figure S1) that engenders a unique suite of physicochemical properties[19,26,27]. The central germanene-like Ge sub-lattice facilitates high electron mobility[19,26], while the encapsulation by heavy Te atoms introduces strong spin-orbit coupling and large anharmonicity[19,27]. Consequently, GaGeTe combines excellent electrical transport with low thermal conductivity, rendering it a competitive candidate for p-type transistors[28], infrared optoelectronics[29], high-efficiency thermoelectrics[19,27,30], and topological spintronics[26,31]. Crucially, this structural hierarchy—characterized by strong intralayer covalent bonding coupled solely by weak interlayer vdW forces along the c-axis—provides the essential physical foundation for the ultra-high plasticity and unique deformation mechanisms explored in this study.

GaGeTe single crystals were synthesized using the modified Bridgman method[28,32]. Due to the peritectic nature of the system, GaTe impurities inevitably coexist within the GaGeTe matrix. To minimize these impurities, we employed a meticulous mechanical separation process to isolate high-quality crystals. Although X-ray diffraction (XRD, sample #3 in Figure S2) confirms surface phase purity by the absence of the characteristic GaTe peak at $2\theta = 24°$, we acknowledge that XRD alone cannot fully rule out internal inclusions. Typically, such inhomogeneities act as stress concentrators that trigger premature brittle fracture.

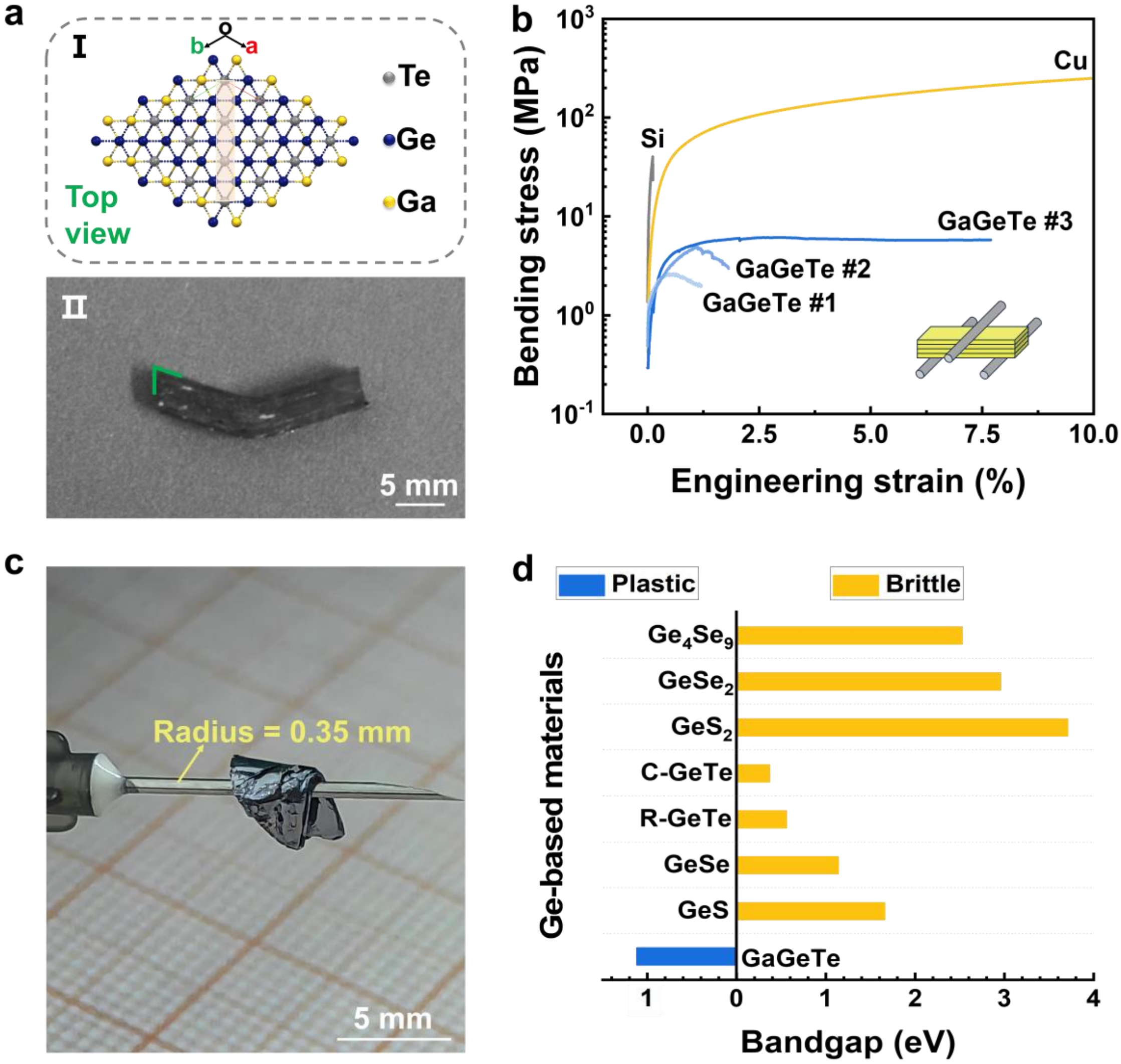

**Figure 1.** Deformability of the GaGeTe crystal. (a) Crystal structure and optical images of GaGeTe. (I) Crystal structure of GaGeTe projected onto the (001) plane. The light pink shaded area indicates the crystal cutting orientation for the three-point bending test. (II) Optical image of the corresponding bulk sample after the test, demonstrating plastic deformation without fracture. (b) Stress–strain curves of the three-point bending test. Typical plastic polycrystalline copper (Cu), brittle Si, and XRD-impure GaGeTe (sample #1 and #2) are shown for comparison. The inset in (b) shows the applied force along the c axis of the crystals. (c) Optical image of a bulk GaGeTe being bent around a syringe needle (radius: 0.35 mm) without fracture. (d) Comparison of the deformability and bandgap of GaGeTe with other Ge-based materials. GaGeTe is the only Ge-based semiconductor material with excellent plastic deformation capability. The bandgap data are collected from reported literature[28,32,41–46]. Note that the values may vary due to different calculation and testing methods; representative data are provided here for general reference.

Our subsequent characterizations (Figures 1b and S2) confirm this failure mechanism in samples with significant segregation. We observed that increasing GaTe content led to premature brittle fracture or inconsistent failure strengths due to phase segregation-induced stress concentrations, despite GaTe itself being a plastic semiconductor[24]. Notably, the XRD phase-pure GaGeTe crystals exhibited considerable tolerance against potential stress inhomogeneities caused by trace inclusions, sustaining a large bending strain of 7.5% without fracture (Figure 1a-II). Furthermore, the material yields at a remarkably low stress of approximately 4.43 MPa—approximately 1/14 of the yield strength of polycrystalline copper (~61.7 MPa). This deformation mechanism is visually evidenced by the acute angles formed at the ends of the bent specimen, indicating extensive interlayer gliding between the van der Waals layers that effectively dissipates applied stress, consistent with deformation modes reported in other layered systems[18]. Moreover, the exceptional macroscopic plasticity of bulk GaGeTe is highlighted by its ability to be wrapped around a syringe needle with a radius of curvature as small as 0.35 mm without fracturing (Figure 1c). Unlike other brittle Ge-based semiconductors (Figure 1d), GaGeTe stands alone as the only material found to date with such extreme plasticity. These findings provide the first systematic mechanical profile of bulk GaGeTe, underscoring its unique position among Ge-based semiconductors.

To investigate the structural origins of the macroscopic plasticity observed in GaGeTe, we conducted a series of multi-scale characterizations. Firstly, scanning electron microscopy (SEM) imaging demonstrates the material's flexibility (**Figure 2**a). An initially flat, layered single crystal was manipulated into a complex "Z"-shaped geometry without fracture. Upon re-flattening, the crystal exhibited permanent creases rather than brittle failure, confirming its plastic nature (original stitched SEM micrographs are provided in Figure S3). The acute-angle morphology formed at the ends of the bent crystal is characteristic of marked interlayer gliding, which aligns with the deformation behaviors observed during macroscopic three-point bending tests. These observations indicate that vdW interlayer gliding is one of the primary deformation mechanisms. To further probe the details of this gliding process, eccentric micro-compression experiments were performed on c-axis oriented micropillars. As shown in Figure 2b (original SEM data are provided in Figure S4), the sidewalls of the compressed GaGeTe micropillar (top diameter ~1.5 μm) exhibit a clear step-like morphology. This distinct morphological feature provides direct microscopic evidence of relative gliding between layers. The schematic diagram of this gliding process is shown in Figure 2c.

We calculated the interlayer slipping energy ($E_s$) for GaGeTe and its binary counterpart rhombohedral GeTe (R-GeTe, hereafter referred to as GeTe), which is brittle, to elucidate the fundamental physical origin of this facile interlayer gliding. As illustrated in Figure 2d, density functional theory (DFT) calculations were performed along the (001)[100] direction for both materials to ensure a rigorous comparison. The results reveal that the interlayer $E_s$ for GaGeTe along this path is extremely low, approximately 0.026 eV/atom. This ultralow energy barrier is a critical factor for its macroscopic softness. Despite the similar composition and layered structure, GeTe exhibits a substantially higher $E_s$ (~0.234 eV/atom) along the same slipping vector. This energy disparity is intrinsically linked to the structural geometry of the van der Waals gap. To understand the pronounced difference in mechanical behavior between the plastic GaGeTe and its brittle analogue GeTe, we analyzed the correlation between slipping energy and interlayer spacing (Figure 2e). GaGeTe possesses a wide interlayer spacing of 3.82 Å, which effectively attenuates the orbital overlap and steric hindrance between adjacent layers, thereby facilitating smooth gliding with minimal energy cost. In sharp contrast, GeTe is characterized by a constricted interlayer spacing of only 2.15 Å. This tight confinement significantly enhances the interlayer interaction, creating a steep energy barrier that locks the atomic layers in place. The brittleness of GeTe is further evidenced by its fracture behavior. Compression experiments reveal a "corrugated paper"-like morphology at the fracture cross-section (Figure S5). This structure hinders further gliding and induces irregular local stress, whereas the expanded interlayer spacing in GaGeTe allows for continuous, planar gliding. Consequently, the combination of a large interlayer spacing and ultralow slipping energy confers unprecedented plastic deformability upon GaGeTe, distinguishing it from conventional germanium-based semiconductors. In comparison of other layered materials (Figure 2f and Table S1), the $E_s$ of GaGeTe is comparable to those of known plastic van der Waals materials (e.g., $SnSe_2$, ~0.022 eV/atom[33]) and even lower than that of typical plastic inorganic semiconductors such as $Ag_2S$ (~0.166 eV/atom[34]). However, while these energetic criteria rationalize the facile interlayer motion, they do not fully elucidate how the intralayer lattice maintains its structural integrity without brittle fracture during such extensive gliding.

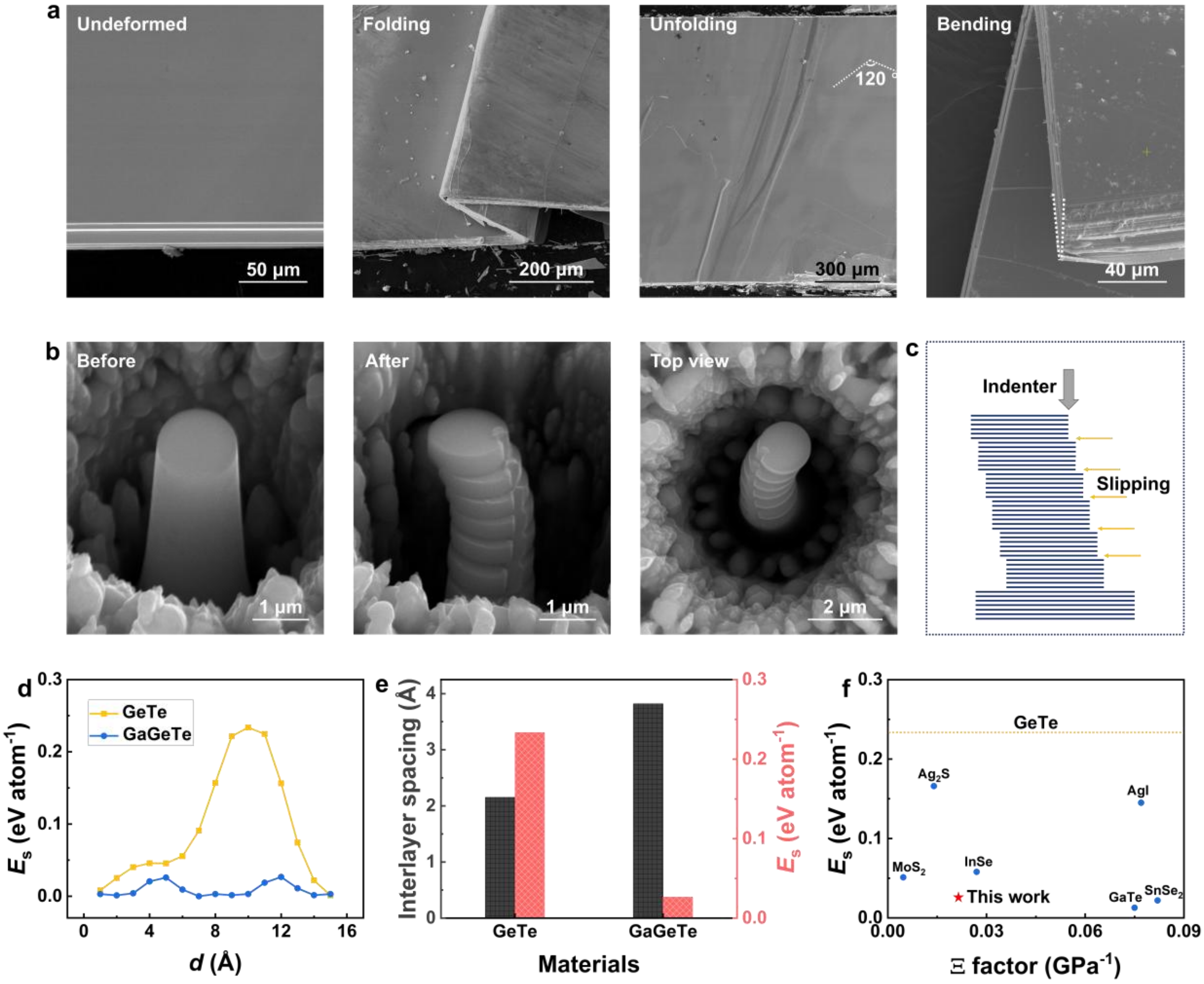

**Figure 2.** Interlayer gliding mechanism and deformability evaluation of GaGeTe. (a) SEM images demonstrating the plasticity of GaGeTe, where an initially flat single crystal subjected to deformation displays permanent creases and characteristic acute-angle edge morphologies, verifying interlayer gliding-mediated plastic deformation. (b) SEM images of a GaGeTe micropillar before and after the eccentric compression test along the c-axis, revealing interlayer gliding. (c) Schematic diagram of the gliding mechanism induced by eccentric compression of a micropillar. (d) Comparison of slipping energies between GaGeTe and GeTe. (e) Statistical analysis of the slipping energy and interplanar spacing of GaGeTe and GeTe. The interplanar spacings were calculated from Crystallographic Information Files (CIF) obtained from the Materials Project database (IDs: mp-8211 and mp-938). (f) Comparison of the deformability factor Ξ and slipping energy of GaGeTe with those of classical plastic van der Waals materials and Ge-based brittle materials (e.g., GeTe, indicated by the yellow dashed line). The factor Ξ was calculated based on the method in Reference 47[47], using cleavage energy data from Figure S6 and elastic stiffness constants from Table S2.

To unravel the atomic-scale origin of this stability, we further investigated the structural evolution of the GaGeTe lattice under deformation using high-angle annular dark-field scanning transmission electron microscopy (HAADF-STEM). A cross-sectional lamella was prepared from a bent region of a GaGeTe single crystal via focused ion beam (FIB) milling (**Figure 3**a). The bright-field (BF) TEM image (Figure 3b) reveals that the deformation is accommodated through the formation of a distinct kink band. Detailed observation shows that stress concentration leads to the formation of local bulges alongside the interlayer gliding. While some regions (Figure 3b-I) undergo smooth gliding, others (Figure 3b-II) develop pronounced kinking-like structures to accommodate the strain. Crucially, the material sustains this complex deformation and maintains a structural integrity throughout the bent region without fracturing.

To verify the internal ordering of the atoms, atomic-resolution HAADF-STEM imaging was performed on both the undeformed matrix (Figure 3c) and the highly stressed kinked region (Figure 3d), corresponding to region I and region II in Figure 3b, respectively. In these images, the contrast of atomic columns follows a $Z^{1.7}$ dependence, rendering the heavier Te atoms ($Z$=52) clearly brighter than the Ga ($Z$=31) and Ge ($Z$=32) atoms. In the deformed region, the atomic arrangement transforms into a curved "racetrack" pattern, characteristic of a crystal lattice undergoing severe bending. The systematic reduction in the number of atomic columns per layer ($\Delta n = 2$) within the bending region is attributed to discrete interlayer gliding, which acts to accommodate the interlayer arc-length mismatch induced by the bending geometry. The relative displacement resulting from this gliding accumulates towards the sample ends, ultimately leading to the acute-angled morphology observed in macroscopic and SEM images.

Notably, the lattice remains devoid of dislocations or other line defects. Extensive atomic-resolution images acquired from various locations and at different magnifications (Supporting Information File 2) consistently substantiate this structural integrity throughout the deformed regions. Furthermore, Selected Area Electron Diffraction (SAED) patterns (Figure 3e) confirm that, despite a bending angle reaching 31° (defined as in Reference), the lattice within the kink band retains a single-crystal structure without undergoing twinning, phase transitions, or amorphization. The absence of such conventional defects indicates that the plasticity is governed solely by interlayer gliding and intralayer distortion.

Quantitative analysis of the lattice metrics reveals substantial anisotropic distortion (Figure 3f). In the stress-concentrated center of the bending, the lattice symmetry deviates from the ideal equilibrium state to accommodate the geometric curvature. Specifically, the intralayer thickness is compressed by tens to hundreds of picometers due to the local bending moment. Conversely, the weak van der Waals interactions allow the interlayer spacing to expand by approximately 80–90 pm, effectively relaxing the mismatch strain between adjacent layers.

A magnified view of the bending center (Figure 3g) presents the fine details of this intralayer distortion. The bending induces an asymmetric strain distribution within individual layers. The upper sub-layers experience compressive stress, resulting in bond contraction, while the lower sub-layers undergo tensile stress, leading to bond elongation. This gradient of bond lengths accumulates along the c-axis, enabling the crystal to sustain large curvature angles.

It is pertinent to note that such atomic-scale lattice distortions are widespread characteristics of layered van der Waals materials; however, prior investigations have primarily identified distortions as secondary effects of native defects (e.g., antisite defects). Consequently, the field has disproportionately emphasized the influence of interlayer interactions and defect dynamics, often obscuring the intrinsic mechanical contribution of the lattice distortion itself and the underlying bonding character changes that define intralayer stability and plasticity. To further conceptualize how these local geometric adjustments prevent fracture, we constructed a simplified ball-and-spring model (Figures 3h, 3i). In this framework, atoms are treated as rigid nodes and chemical bonds as flexible springs capable of non-linear deformation. Importantly, the model reveals that the GaGeTe lattice accommodates strain through bond rotation and angle variation, rather than merely relying on linear bond stretching or compression. This mechanism is directly supported by our atomic-resolution HAADF-STEM observations (Figure S7), where quantitative analysis identifies distinct bond angle deviations from the equilibrium state, providing the physical basis for the flexible rotation capability. Under vertical loading, while the "springs" on the convex and concave sides undergo respective stretching and compression, the bond angles simultaneously adjust to redistribute strain energy. Consequently, this synergistic interaction—combining interlayer gliding with intralayer lattice distortion—effectively dissipates local stress. This critical atomic-level response complements the interlayer gliding mechanism, explaining how GaGeTe stabilizes kink band formation and absorbs mechanical energy plastically without severing its bonding network.

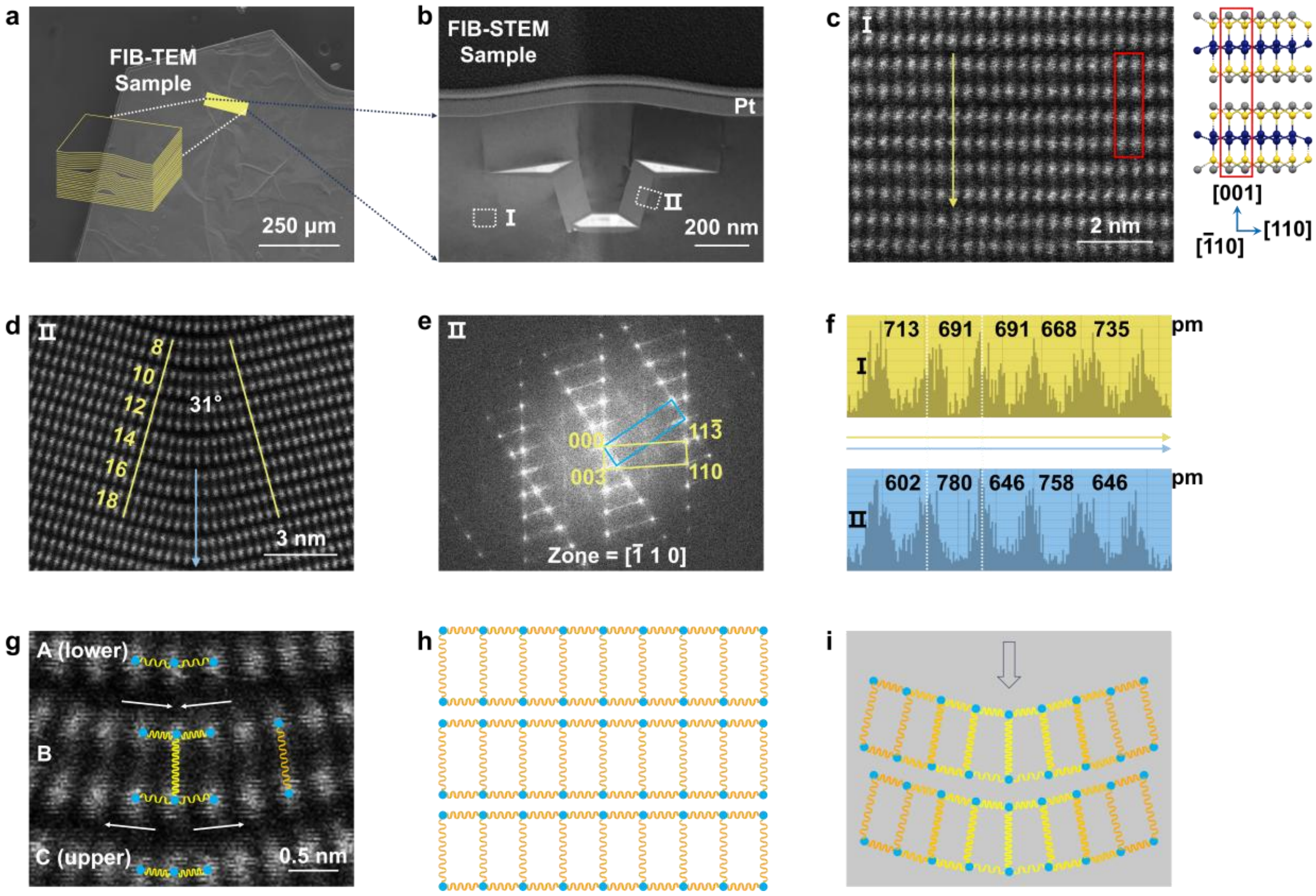

**Figure 3.** Atomic-scale investigation of intralayer lattice distortion in GaGeTe upon bending. (a) SEM image showing the location where the TEM lamella was prepared by FIB (at a bent region in the GaGeTe sample). (b) Bright-field (BF) TEM image of the cross-sectional view of the bent GaGeTe lamella. (c) High-angle annular dark-field scanning transmission electron microscopy (HAADF-STEM) image of the undeformed matrix (labeled I) in (b). (d and e) HAADF-STEM image and selected area electron diffraction (SAED) pattern taken from the bent region (labeled II) in (b), respectively. (f) Comparison of the measured interplanar spacings in the undeformed and bent regions. The arrows in (c) and (d) indicate the measurement locations. (g) Magnified view of the central area in the HAADF-STEM image in (d). (h and i) Schematic illustrations of the simplified spring models for the undeformed and bent GaGeTe lamella, respectively.

Based on the systematic investigation of the macroscopic plasticity, microscopic gliding, and atomic-scale lattice distortion of GaGeTe, we employed Density Functional Theory (DFT) calculations to elucidate the chemical origins of the distinct mechanical contrast—plasticity versus brittleness—between GaGeTe and GeTe. Specifically, we analyzed the Crystal Orbital Hamilton Population (COHP) to evaluate bond stability under strain and employed the Electron Localization Function (ELF) to characterize the intrinsic bonding nature. Since the preservation of intralayer structural integrity is a prerequisite for interlayer gliding, i.e., the lattice must withstand distortion without bond rupture to enable gliding, quantifying bond strength under deformation is critical. To facilitate the application of uniaxial tensile strain along specific crystallographic directions and ensure computational consistency, the hexagonal unit cells of both materials were transformed into orthogonal supercells (**Figure 4**a), defined such that the orthorhombic axes, $a_{ortho}$ and $b_{ortho}$, correspond to the hexagonal [100] and [120] directions, respectively. Within this setup, we systematically computed the integrated COHP (-ICOHP) for the constituent bonds in both systems under equilibrium conditions and varying tensile strains (5% and 10%). These bonds include Ge-Te in GeTe, and Ga-Te, Ga-Ge, and Ge-Ge in GaGeTe, with their respective COHP plots shown in Figures S8 and S9.

The value of -ICOHP serves as an effective indicator of the strength of chemical bonding interactions. The application of tensile strain inevitably induces geometric distortions, specifically characterized by bond elongation and angular deviations (Table S3), which results in a natural decrease in -ICOHP values for all bonds as illustrated in Figure 4b. However, the magnitude of this reduction reveals a fundamental difference between the two materials. In GeTe, the crystal structure is governed by intrinsic metavalent bonding, a state sharing about one electron between adjacent atoms that features partial electron delocalization, rendering the lattice susceptible to Peierls distortion and collective bond breaking under stress. This manifestly low bonding strength, a hallmark of the metavalent bonding family, underscores the material's limited resistance to mechanical failure. This is evidenced by the lack of bonding states below the Fermi level in the Ge-Te COHP curves (Figure 4c), yielding a notably low -ICOHP of only 1.32 eV at the unstrained state. Conversely, GaGeTe is stabilized by a significantly more robust bonding network underpinned by distorted $sp^3$ hybridization, characterized by bond angles deviating from the ideal 109.5° (Table S3). Within this environment, Ge–Ge interactions exhibit strong bonding features reminiscent of germanene, ensuring remarkable cohesion: even under a simulated 10% tensile strain, the -ICOHP values for Ga–Te, Ga–Ge, and Ge–Ge bonds all remain above 3.0 eV. To corroborate the robust nature of this supporting network, Electron Localization Function (ELF) analysis was performed on GaGeTe (Figure S10). The results yield high localization values for the backbone constituents: ~0.75 for Ge-Ge bonds and ~0.73 for Ga-Ge bonds. These values unequivocally confirm the formation of a covalent structure, distinct from the delocalized metavalent bonding in brittle GeTe. This robust covalency is the fundamental enabler of plasticity: instead of succumbing to intralayer fracture, the high-strength backbone maintains strong cohesion even during severe lattice distortion, thereby allowing the accumulation of the strain necessary to drive macroscopic plasticity via interlayer gliding.

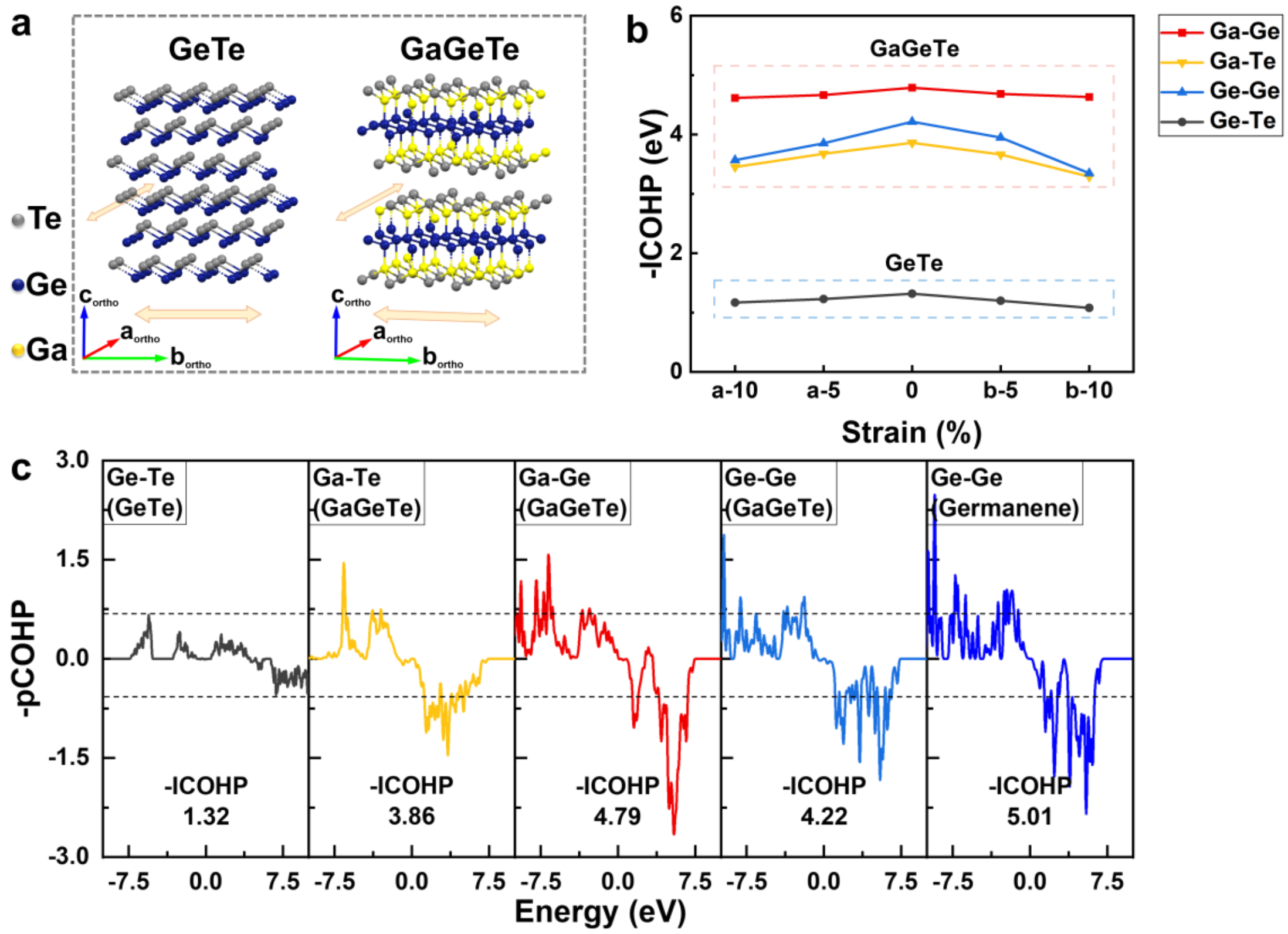

**Figure 4.** Chemical bonding origin of the plasticity in GaGeTe. (a) Comparison of the crystal structures of GeTe and GaGeTe modeled as orthogonal supercells for tensile strain calculations. (b) Variation of the -ICOHP values for different chemical bonds (the Ge–Te bond in GeTe; the Ga–Te, Ga–Ge, and Ge–Ge bonds in GaGeTe; and the Ge–Ge bond in germanene) as a function of uniaxial tensile strain along the $a_{ortho}$- and $b_{ortho}$-axes. (c) Crystal Orbital Hamilton Population (COHP) curves for the corresponding bonds, highlighting the significantly stronger bonding interactions in GaGeTe compared to the weaker bonding in GeTe.

## 3. Conclusion

Our study underpins GaGeTe as the first germanium-based inorganic semiconductor to exhibit ultra-high plasticity, thereby addressing the long-standing brittleness issue in Ge-based chalcogenides. By integrating multi-scale characterizations ranging from macroscopic characterization to atomic-resolution AC-TEM imaging and first-principles calculations, we have elucidated a unique deformation mechanism co-dominated by interlayer gliding and intralayer lattice distortion. This mechanical flexibility is structurally originated from a wide van der Waals gap, which lowers the energy barrier for interlayer gliding (0.026 eV/atom), and a unique sextuple-layer structure that accommodates severe bending. Crucially, our chemical bonding analysis reveals that the origin of this plasticity lies in the specific bond engineering of the ternary system: GaGeTe avoids the formation of half-bond order metavalent Ge-Te bonds characteristic of its brittle analogue GeTe. Instead, it constructs a robust bonding network comprised of strong covalent Ga-Te, Ga-Ge, and Ge-Ge bonds that maintain high bond strength even under large tensile strains. These findings not only offer deep insights into the physical mechanisms and chemical origins of plasticity in layered inorganic materials but also establish a "bond engineering" strategy for the rational design of next-generation plastic semiconductors for flexible electronics.

## 4. Experimental Methods

*Growth of single-crystal GaGeTe:*

High-purity Ga (5N, 99.999%), Ge (5N, 99.999%), and Te (5N, 99.999%) were obtained from ZhongNuo Advanced Material (Beijing) Technology Co., Ltd., and used as raw materials. Stoichiometric amounts of Ga, Ge, and Te (with a molar ratio of Ga : Ge : Te = 1 : 1 : 1) were loaded into a carbon-coated quartz ampoule under vacuum, which was then flame-sealed. The synthesis was carried out in a programmable tube furnace using the following thermal protocol: the temperature was first raised to 950 ℃ at a rate of 10 ℃/min and held for 12 hours for homogenization; then slowly increased to 980 ℃ at 1 ℃/min and maintained at this temperature for 3 days to ensure complete reaction. Subsequently, the melt was cooled to 850 ℃ at 5 ℃/min. The crystallization process was then performed by slowly cooling the ampoule from 850 ℃ down to 750 ℃ at a constant rate of 2 ℃/h. Finally, the furnace was switched off, and the sample was allowed to cool naturally to room temperature.

*Growth of single-crystal GeTe:*

The synthesis of GeTe was performed using a modified melting-annealing method adapted from previous reports[35,36]. High-purity Ge (5N, 99.999%) and Te (5N, 99.999%) were obtained from ZhongNuo Advanced Material (Beijing) Technology Co., Ltd., and used as raw materials. Stoichiometric amounts of Ge and Te were loaded into a carbon-coated quartz ampoule under vacuum, which was then flame-sealed. The ampoule was placed in a tube furnace, heated to 980 ℃, and maintained at this temperature for 6 h to ensure complete melting and homogenization of the reaction melt. Subsequently, the melt was cooled to 750 ℃ at a rate of 10 ℃/min. The temperature was then slowly lowered from 750 ℃ to 700 ℃ over a period of 3 days. Following this, the sample was annealed at 700 ℃ for another 3 days to promote phase formation and homogeneity. Finally, the furnace was switched off, and the obtained sample was allowed to cool naturally to room temperature, yielding the rhombohedral GeTe phase.

*Characterization:*

Three-point bending engineering stress−strain tests were performed using an Instron® 5566 universal testing machine with a constant loading rate of 0.05 mm $min^{-1}$. The bulk GaGeTe specimen had dimensions of 4.24 × 4.24 × 20.20 $mm^3$. The scanning electron microscope (SEM) images were carried out on Thermo Scientific Scios 2 DualBeam. The transmission electron microscopy (TEM) and spherical aberration corrected transmission electron microscopy (AC-TEM) data of GaGeTe were obtained using the Talos F200X and JEM-ARM200F NEOARM. The TEM samples were done by a focused ion beam (FIB, FEI Helios NanoLab 600i DualBeam). Optical photographs were taken by VIVO IQOO Z7I. The GaTe crystal was examined by X-ray diffraction using Cu Ka Source (Panalytical Instruments, X 'PERT). Nanoindentation data were collected on an Agilent G200.

*DFT computational details:*

In this work, all calculations were performed using the plane-wave code Quantum Espresso.[37] The Perdew-Burke-Ernzerhof functional within the generalized gradient approximation was used to describe the electron exchange-correlation interactions.[38] The Van der Waals interaction was handled by semiempirical correction (DFT-D2).[39] The kinetic energy cutoffs for wavefunctions and charge density were 80 and 400 Ry, respectively. The convergence criteria for electronic structure iterations and the force on each atom during structure relaxation were set to $10^{-9}$ Ry and $10^{-4}$ Ry/Bohr, respectively. We sampled the Brillouin zone with a 5×5×1 Monkhorst-Pack k-points for geometry optimizations and electronic structural calculations. A vacuum space of 30 Å was used to avoid interactions between the periodic images of layers in the z direction. The crystal orbital Hamilton population (COHP) analysis was carried out using the LOBSTER code.[40]

*Declaration:*

During the preparation of this manuscript, the authors used Gemini solely for the purpose of improving language fluency and grammar in the initial draft. All scientific content, data analysis, interpretation, and conclusions are entirely the product of the authors' work. The AI tool was not used to generate any scientific ideas, data, or interpretations, and it is not listed as an author.

**Conflict of Interest**

The authors declare no conflict of interest.